# Electron Attachment to DNA and RNA Nucleobases: An EOMCC Investigation

**Achintya Kumar Dutta*, Turbasu Sengupta, Nayana Vaval and Sourav Pal**

Physical Chemistry Division, CSIR-National Chemical Laboratory, Pune-411008, India

**Abstract:** We report a benchmark theoretical investigation of both adiabatic and vertical electron affinities of five DNA and RNA nucleobases: adenine, guanine, cytosine, thymine and uracil using state-of-the-art equation of motion coupled cluster (EOMCC) method. We have calculated the vertical electron affinity values of first five electron attached states of the DNA and RNA nucleobases and only the first electron attached state is found to be energetically accessible in gas phase. An analysis of the natural orbitals shows that the first electron attached states of uracil and thymine are valence-bound type and undergo significant structural changes on attachment of excess electron, which is reflected in the deviation of the adiabatic electron affinity from the vertical one. On the other hand, the first electron attached state of cytosine, adenine and guanine are dipole-bound type and their structure remain unaffected on attachment of an extra electron, which results in small deviation of adiabatic electron affinity from that of the vertical one. Vertical and adiabatic electron affinity values of all the DNA and RNA nucleobases are negative implying that the first electron attached state are not stable, but rather resonance states. Previously, reported theoretical studies had shown scattered results for electron affinities of DNA and RNA bases, with large deviations from experimental values. Our EOMCC computed values are in very good agreement with experimental values and can be used as a reliable benchmark for calibrating new theoretical methods.

*ak.dutta @ncl.res.in

# 1. INTRODUCTION

The electron affinities (EA) are among the most important intrinsic properties of DNA and RNA nucleic acid bases (NAB) having both theoretical and experimental interest. Accurate determination of EA of NAB essential for the understanding of a variety of phenomena related to the formation of transient charged radicals[1-4] within the DNA and RNA strands, which are responsible for various radiation-induced phenomena, such as, radiation response of genetic materials[5-9], interaction of DNA with proteins[10, 11], alteration and deletion of genetic information causing cancer (genome mutagenesis)[12, 13]. In spite of its vast importance in different fields, including biochemical and medical based technologies[14, 15], very few reliable experimental data are available and the results are scattered within a broad range from negative to positive[16] values. The main reason behind the experimental uncertainty is the presence of different tautomers of NAB in the gas phase, which are very close-lying in energy. Even for the same tautomer, there exist two different possibilities[17-20] of a negatively charged system. First one is the conventional valence bound (VB) structure, containing an extra electron in the anti-bonding molecular orbital, which leads to the significant structural changes in the species relative to its neutral precursor. There is a second possibility in highly polar molecules[21-23] (dipole moment equal or higher[24] than 2.5 D), where the extra electron is weakly bound to molecule by means of electrostatic charge-dipole interactions (DB), and consequently the molecular structure remains mostly unchanged from that of the neutral precursor. Now both VB and DB anion of NAB can remain in a very close range in energy and different experimental conditions may lead to preferential condition of one or the other, which introduces large error bars in experimentally measured electron affinities of NAB.

Theoretical determination of EA of NAB is also very difficult, however, for completely different reasons. The density functional theory (DFT) calculations show high dependence[25-28] of the results on exchange-correlation functional used in the calculations. The state-of-the-art *ab-initio* quantum-mechanical calculations, although being more accurate, are difficult to perform because of the computational constraints, which arise due to the use of highly diffuse basis set having maximum radial and angular flexibility required to model weakly bound electron in the dipole bound (DB) NAB anions. Considering the whole scenario, the accurate theoretical estimation of EA of DNA and RNA nucleobases is one of the prime fields of work for theoretician since last decade. Different theoretical methods, starting from semi-empirical methods[29, 30], qualitative Hartree-Fock Koopmans' method[31] to MP2[16, 31], MP4[16] or multi-configurational CASPT2[16] perturbation treatment, density functional theory (DFT)[25-28] with different functionals, and even state-of-the-art coupled cluster method[16, 32] have been used for theoretical estimation of EA values in NAB. All of these methods are employed using a variety of basis sets. In spite of all these efforts made, most of the theoretically estimated EA values are out of the range compared to the experimental values concerned, and like the experimental results, are scattered. Consequently, it is difficult to draw a clear conclusion about the exact EA values of NAB from theoretical data available in literature and it calls for a new study using an accurate electron correlation method along with a sufficiently large basis set augmented with diffuse functions for accurate estimation of EA of nucleobases.

The single reference coupled cluster method is well known for the systematic incorporation of dynamic correlation effects and has been used for the energy[33, 34], geometry[35-37] and IR frequencies[38-40] of closed shell molecules with great success. Although single reference coupled cluster includes the dynamic correlation in a systematic way, it fails to account for the non-dynamic correlation prevailing in quasi-degenerate situations, which occurs in bond stretching, open shell radicals and molecular excited states. A multi-reference coupled cluster (MRCC) method[41-57] can describe dynamic and non-dynamic correlation in a balanced way. However, like any other multi-reference method, MRCC is conceptually difficult and hence the use of the method in the elucidation of chemical problems, require considerable experience and expertise. Parallel to the MRCC methods, the equation of motion coupled cluster (EOMCC) method[58-62] provides a black-box tool for a balanced description of dynamic and non-dynamic correlation. The EOMCC method has been successfully used to compute the accurate energy difference, such as ionization potentials (IP)[59], electron affinities (EA)[58] and excitation energies (EE)[60-62].

Krylov and co-workers[63-66] have used EOMIP-CCSD method for accurate prediction of ionization energies of NAB and obtained very good agreement with experimental values. However, to the best of our knowledge, no such studies have been performed for calculating electrons affinities of NAB. There are obvious difficulties associated with the use of EOMEA-CCSD to investigate EA of large molecules, such as DNA bases. For the accurate prediction of EA, one need to use large basis set with diffuse functions and this makes calculation of EOMEA-CCSD very expensive and unlike EOMIP-CCSD method, EOMEA involves 4-particle intermediates, which has large storage requirement and slows down the calculations significantly because of disk I/O. Very recently, an lower scaling and 4-particle intermediates free approximation to EOMEA-CCSD developed, which can be applied to large systems[67].

The aim of this paper to obtain benchmark values of vertical and adiabatic electron affinities of DNA base pairs using EOMEA-CCSD method in a sufficiently large basis set, which can be used to calibrate new theoretical methods and discard the less reliable experimental data. The paper is organized as follows: The next section contains the computational details of the calculations. Results and discussions are followed in section 3. Section 4 contains the concluding remarks.

## 2. COMPUTATIONAL DETAILS:

All the structures presented in this paper were optimized in DFT/aug-cc-pVTZ level of theory with double hybrid B2PLYP functional. The double-hybrid B2PLYP functional has been shown to produce geometric parameters with accuracy comparable to the CCSD(T) method, for both neutral molecules and radicals[68]. The Vertical and adiabatic electron affinities were calculated using EOMEA-CCSD method in aug-cc-pVDZ and aug-cc-pVTZ basis sets[69, 70]. Gaussian09[71] was used for DFT calculations. CFOUR[72] has been used to perform all the coupled cluster calculations. One diffuse f function is removed from the aug-cc-pVTZ basis set to keep the problem computationally viable.

## 3. RESULTS AND DISCUSSION

### A. Comparison of Ground and Anionic state geometries

An analysis of geometries of neutral and anionic nucleobases shows considerable deviation in bond length, bond angle, and dihedral angle between neutral species and the anions. Table 1 shows maximum deviation in structural parameters in the process of formation of anions from neutral geometries. Rightmost column of the table contains the root mean square deviation value (RMSD) between neutral and anionic structures. Figure 1 shows the molecular structures of all five nucleobases with maximum deviations as well as other major structural changes, in each anion.

The maximum deviation in bond lengths and bond angles between neutral and anionic state geometries, among all NAB, are seen in the case of uracil (0.07Å and $5.51^0$ respectively). Whereas, the maximum change in dihedral angle is observed in case of thymine $-67.4^0$, which is due to the twisting of methyl ($-CH_3$) side chain in thymine anion. The pyrimidine NAB, uracil and thymine are structurally similar, e.g. both have two carbonyl groups in the ring, meta to each other. Therefore, they undergo similar structural changes upon electron attachment, for example, each of them shows an almost equal amount of maximum bond length (+0.070 Å and +0.063 Å respectively) and bond angle (-5.51$^0$ and -4.93$^0$ respectively) deviation from neutral to anionic species. The structural changes of uracil and thymine can be better understood from the analysis of HOMO of the neutral and SOMO of the anionic species, highlighted in Figure 2. In both cases, the singly occupied molecular orbital is localized; indicating the formation of valence bound anions. Strong bonding interaction between C1-C2 in the neutral uracil is replaced by antibonding interaction in the anion. Consequently, the C1-C2 bond of the anion gets elongated from that in neutral uracil. The associated bond angle N1-C1-C2 gets distorted by about 5.51$^0$ and the dihedral angle (C2-C1-N1-C4) is changed by 16.16$^0$. Similarly, in thymine, a significant anti-bonding interaction is introduced between C2-C3 (+0.063 Å) in the anion, in place of bonding interaction present in the neutral state, resulting in the distortion of the associated N2-C3-C2 angle by 4.93$^0$. The high value of angle decrement and dihedral changes in cases of uracil, as well as thymine indicates that certainly there is an angle strain induced ring distortion in the anions. Ring puckering in the optimised geometries of uracil and thymine are clearly observed in the planer view of uracil and thymine anions. Figure 2 shows that geometries of both the anions are deviated from the planarity, and the farthest hydrogen is shifting outwards from the ring as a consequence of ring puckering (H1 in case of uracil and H4 in thymine). Among the other mentionable distortion of uracil anion, both N1-C1 and C2-C3 bonds are distorted by an amount 0.05Å and angle N2-C3-O1 by 4.4$^0$. The ring distortion in uracil and thymine leads to higher RMSD value than rest of the NAB, which is prominent in the RMSD plot in Figure 3. The unusually high RMSD value in case of thymine is, firstly, due to the twisting of the methyl group in its anion. Secondly, thymine shows a large deviation of C2-C3 bond length (+0.063Å) and N2-C3-C2 bond angle (-4.93$^0$) respectively. Both of these can explain why thymine anion has the highest RMSD value among all the nucleobases. Figure 4 shows that the SOMO of other three nucleobases, adenine, guanine and cytosine, are very diffuse in nature, indicating the formation of the dipole-bound anions, from which it can be expected that these three NAB would show small structural deviations in the anionic state. Adenine has the lowest RMSD value, among all the nucleobases. The major changes in geometrical parameters observed in adenine are twisting of amino group of about 7.36$^0$ around the ring, increment of C1-N5 bond by ~0.017Å and decrement of H2-N5-H3 angle by 2.3 degree. Twisting of the amino group, similar to adenine, is observed in the case of guanine also, a 9.76$^0$ rotation of NH2 group is noticed. However, the corresponding C-N bond behaves differently to electron attachment. In case of adenine, the corresponding N-C bond (N5-C1) stretches by 0.017Å, but in guanine similar bond (N5-C5) gets shorten by 0.017Å, compared to its neutral precursor. The other important changes in geometrical parameters of guanine anion are increments of angle H3-N5-C5 ($2.24^0$), H1-N3-C4 ($\sim 1^0$) and of bond C5-N4 (0.01Å). The shortening of N3-C4 bond by 0.01Å and C2-C4-O1 angle by $0.9^0$ is also noticed. No ring distortion is observed in guanine anion. The RMSD value of cytosine is highest (0.067) among the dipole bound anions, which may seem unusual at first look, but a careful inspection reveals that the only major contributor to the high RMSD value is the huge twisting of $NH_2$ group ($\sim 21^0$). Structural changes in all other parameters are minimal, maximum bond length change is about 0.01 Å (C2-O1) and maximum bond angle change is less than $1^0$(H2-N3-C1).As a matter of fact cytosine shows minimum changes in bond lengths and bond angles upon electron attachment among all the NAB, which is expected from the dipole bound nature of its anion. The ring planarity is preserved in the case of cytosine as well.

**B. Vertical electron affinities of nucleobases:**

Experimental electron affinity values of nucleobases are reported in Table 2. In initial experimental approaches, determination of electron affinity of compounds was mostly based on two fundamental methods (i) photoelectron spectroscopy (PES)[73, 74] and (ii) Rydberg electron transfer (RET)[19, 75, 76]. However, the most modern experimental approach is the electron transmission spectrosopy (ETS)[77], which is specifically suitable for molecule with negative EA. This specroscopic technique calculates the EA of molecules via the detection of transient resonance states of approximately femtosecond in the lifetime.

First reported value of VEAs of valence bound states of nucleobases via ETS method was done by Burrow and coworkers[78]. Later, Desfrancois and coworkers[79] calculated VEA values of nucleobases by using a combination of cluster solvation method and RET spectroscopy. They have predicted uracil and thymine both have vertical EA value of -0.30 eV, whereas the VEA of adenine and cytosine are −0.45 eV and -0.55 eV respectively. Further experimental estimation on VEA of DNA and RNA nucleobases includes enthalpy of formation technique by Harinipriya and Sangaranarayanan[80]. They have predicted VEA value of uracil and thymine as -0.24 eV and -0.53 eV. On the other hand, adenine, cytosine and guanine have been predicted to show VEA values of -0.40 eV, -0.56 eV and -0.79 eV, respectively and these are slightly on the higher side compared to the previous two reports[78, 79].

From the experimental results discussed above, there are few fundamental points about experimental VEA of NAB that should be noted. Firstly, in all the experimental methods, VEA values of all nucleobases are negative. Secondly, experimental VEA values for guanine are absent in most results except for heat of formation method[80]. It is due to the instability of keto form of guanine in gas phase. Guanine has a tendency to isomerise or decompose before reaching the minimum vapour pressure needed for experimentation[79]. Finally, the experimental trend of EA in most sophisticated ETS[78] is U>T>C>A. In Table 3, we have tabulated the computed VEA values obtained in earlier theoretical investigations and our present EOMEA-CCSD results for the first vertical electron attached state of each nucleobases.

An analysis of the theoretical methodologies employed for the calculation of vertical EA of NABs reveals that the VEA values obtained in different *ab-initio* methods differ widely from one another in magnitude of electron affinity. Sometimes even the sign varies depending upon the level of theory employed. The simplest possible approach, which is scaled Koopmans' approach[31] in D95V set, gives values very close to the experimental range. However, Staley and strand[81] have shown that this agreement is due to a fortuitous error cancellation between a bad basis set and the inappropriate method, and the agreement deteriorates in larger basis sets. The DFT methods[25-28] show mostly scattered values ranging from high negative to small positive value, depending upon the functional and basis set used. The MP2 methods[16] lead to very high negative values in 6-31G(d) basis set. However, the values become less negative with the use of diffuse aug-cc-pVDZ. Spin contamination also has a significant effect on electron affinities calculated in MP2 method. The use of projected MP2 method makes the predicted electron affinity values less negative i.e., closer to the experimental value. The CCSD(T)[16] and CASPT2[16] give similar values and both of them underestimate (i.e., give more negative value) compared to the experimental values.

Our EOMEA-CCSD calculation shows the best agreement with the experimental values for all five NAB. However, the vertical electron affinity values obtained in EOMEA-CCSD method differ considerably from the previous theoretical reports. The electron affinity in EOMEA-CCSD method is much higher than the MP2 and CCSD(T) values reported by Serrano-Andres[16]. Especially, the vertical EA for uracil is only -0.20 eV in EOMEA-CCSD/aug-cc-pVTZ level of theory, which is much smaller than that reported in the previous theoretical investigation. However, it should be noted that the above CCSD(T) calculations were performed in small aug-cc-pVDZ basis set. On the other hand, Urban and co-workers[32] have reported an electron affinity of -0.15 eV from the OVOS-CCSD(T) calculation in aug-cc-pVTZ basis, which is in the same range of our EOMEA-CCSD value of 0.20 eV. It is also interesting to note that the purine NAB, adenine and guanine, show significantly higher values of VEA in EOMEA-CCSD and the values are in the similar range of the pyrimidine NAB, in contrary to previous theoretical reports, where pyrimidine NAB are shown to have much higher VEA value compared to the purine NAB. Especially, guanine, which possesses the lowest electron affinity value in all the previous theoretical methods, shows the highest electron affinity in EOMEA-CCSD method. The lack of reliable experimental values makes it difficult to conclude the accuracy of the VEA value obtained for guanine in EOMEA-CCSD method. However, the EOMEA-CCSD method, which is in best agreement with experiment for all the other four NAB, is expected to perform equally good in case of guanine as well.

The loosely bound electron in NAB anions, may result into multiple quasi-degenerate configurations, thus, their accurate description demands for systematic inclusion of dynamic and non-dynamic correlation, which CCSD(T) fails to include in a balanced way. The CASPT2 results of Serrano-Andrés and co-workers[16] show that inclusion of non-dynamic correlation leads to increase of electron affinity values compared to single reference method. However, the CASPT2 values depend upon the choice of active space, which requires experience and expertise. The EOM based methods, on the other hand, provide a 'black box' way to provide balanced description of both dynamic and non-dynamic correlation and can calculate electron affinity values corresponding to multiple states in a single calculation.

Table 3 shows that the basis set has a significant effect on the EOMEA-CCSD calculated electron affinity values of NAB. On improving the basis set, the VEA values of NAB increase (i.e. become less negative) which signify that the electron attached states of NAB become energetically more stable on improving the basis set. The EOMEA-CCSD values of first valence VEA in both basis show close agreement with the experimental range[78-80]. A quick look upon Figure 5 will clarify the argument. For example, in uracil, EOMEA-CCSD/aug-cc-pVDZ and EOMEA-CCSD/aug-cc-pVTZ values for VEA for the first state are -0.23 and -0.20 eV, respectively, whereas the ETS value[78] is -0.22 eV and enthalpy mesurement[80] shows a VEA of -0.24 eV. Similarly, the first VEA value of Thymine -0.24 eV (obtained in EOMEA-CCSD/aug-cc-pVTZ ) is very close to the experimental ETS[78] (-0.29 eV) and cluster solvation method[79] (-0.30 eV). The VEA values of two other nucleobases, cytosine and adenine also show similar proximity with experimental results. The cytosine's VEA value in EOMEA-CCSD method in both the basis sets (-0.27ev and -0.24 eV in aug-cc-pVDZ and aug-cc-pVTZ basis set, respectively) is closer with ETS measurements[78] (-0.32 eV) and enthalpy of formation[80] data(-

0.40 eV), whereas the VEA value of adenine in EOMEA-CCSD method shows proximity (-0.44 and -0.40 eV in aug-cc-pVDZ and aug-cc-pVTZ respectively) with cluster solvation method[79](-0.45 eV) data. The results obtained in other two experimental methods, i.e., ETS[78] (-0.54 eV) and cluster solvation method[79] (-0.56 eV) for adenine are also in the similar range and just around 0.1 eV below than the EOMEA-CCSD result.

Now the situation is little bit complicated for cytosine and guanine due to presence of multiple possible isomers. The structure of various possible isomers cytosine and guanine are presented in figure 6.

The first electron attached states of different isomers of cytosine are presented in table 4. It can be seen that the C1 form shows the highest, i.e. least negative electron affinity (-0.27 eV in aug-cc-pVTZ). Although, in gas phase the lowest energy isomer is C2b, which shows lowest electron affinity of -0.38 eV. However, the C2b, C1 and C3a isomer is near degenerate in CCSD/aug-cc-pVTZ level of theory. The C3a isomer gives nearly identical electron affinity (-0.25 eV) as that of C1. The C2a and C3b isomer is slightly higher in energy (0.77 and 1.84 kcal/mol,respectively) and shows electron affinity value -0.34 and -0.32 eV, respectively in aug-cc-pVTZ basis set.

Table 5 presents the electron affinity vale corresponding first electron attached states of different isomers of guanine. It can be seen that the G9k form gives the highest electron affinity of -0.19 eV in aug-cc-pVTZ basis set. However, the isomer is near-degenerate with the lowest energy G7k isomer, which shows the electron affinity value of -0.23 eV. The G9Es isomer is also near-degenerate with the G9k and G9es isomer and the former shows the lowest electron affinity value of -0.43 eV in EOMEA-CCSD/aug-cc-pVTZ level of theory. The G9Ea and G7Es isomer is 0.64 ang 3.68 kcal/mol higher in energy than the lowest energy G7k energy and the former two shows electron affinity of -0.37 and -0.35 eV. For rest of the manuscript only isomer having highest electron affinity of cytosine(C1) and guanine(G9k) have been considered.

One of the attactive feature of EOMCC is that it can calculate electron affinity of correspondin to multiple states in a single calculation. Table 4 compiles the VEA values of the first five states of purine and pirimidine NAB computed in EOMEA-CCSD method and Dunning's correlation-consistent aug-cc-pVXZ (X=D,T) basis[69, 70]. The higher electron attached states of different isomers of cytosine and guanine are preseneted in supporting information. A close look at the results in Table 4 shows that only the first vertical electron affinity (state 0 in the Table 4) of NAB can be acesseble in gas phase. The second and higher electron attached states are too high energy to be stable in gas phase.However, the previous theoretical studies on the solvent effects on electron affinity values of NAB have shown[82-84] that the presence of solvent molecules leads to the stabilization of electron attached states, which results in lowering of their energy. Therefore, the second and higher electron affinities of NABs can be accessibile in solvent phase. Here, it should be noted that the calculation of second or higher electron affinites are feasible only in EOM based and other similar difference energy based methods are difficult to be computed in the $\Delta$ based methods used in previously reported theoretical studies. However, the detailed EOMCC study of the solvation effect on first and higher electron attached states is out of the scope of the present study and will be studied in subsequent papers. Here it should be noted that the lifetime of the temporay bound NAB anion depends upon the hight and wildth of the centrifugal potential barier, which is govorned by a complex interplay of short range and long range interaction and needs very careful analysis before reaching any firm consclusions.

Reprising the experimental trend[78-80] of VEA discussed previously, a close inspection of our EOMEA-CCSD result shows the following trend for the first vertical electron affinity in both aug-cc-pVDZ and aug-cc-pVTZ basis set.

$$U>T\sim C>A$$

The trend is same as that of the experiential results, except the fact that thymine and cytosine have near identical VEA. Including guanine in the picture, the overall trend for VEA in EOMCC calculation is found to be $G\geq U>T\sim C>A$ for state 0. However, this trend is not conserved in higher electron attached states, as shown in Table 7.

**C. Adiabatic electron affinities of nucleobases:**

The sign and the magnitude of the adiabatic electron affinities (AEA) of purine and pyrimidine nucleobases are still a matter of debate[85]. Not only the magnitude of AEA varies widely with the nature of theoretical or experimental method applied, but also there is severe confusion about the sign of AEA of nucleobases. Table 8 presents some results of initial experimental approach to evaluate AEA of nucleobases. The main reason of experimental confusion is due to the absence of any direct procedure to measure AEA of molecules. Initially, the AEA of uracil and thymine were determined by the method of substitution and replacement[86], in which both these anions are found to be strongly bound and have an AEA value of 0.75 eV for uracil and 0.65 eV for thymine. Guanine was proven exceptionally stable with an AEA value greater than 1 eV. Later work of Chen and Wentworth[87, 88] using cyclic voltammetry in DMSO solvent (using scaling factor of known EA of acridine and anthracene) proved that uracil & Thymine both have positive AEA of value 0.80 & 0.79 eV respectively. Their results were further supported by semi-empirical multi-configurational AM1-MCCI calculations[29, 30]. On the other hand, Weinkauf and co-workers[89], from their experimental study involving linear extrapolation method, have found uracil with an AEA value of 0.15 eV and thymine with an AEA value of 0.12 eV. However, they have reported an uncertainty of ±0.12 eV in each case. The scattered values and large uncertainty involved with the experimental results make it difficult to arrive at any solid conclusion.

Table 9 shows that in most of the *ab-initio* results of AEA are negative, in contradiction to the experimental results. The DFT-B3LYP[27, 90-95] method shows scattered results, especially in case of uracil and thymine, where the AEA values vary from highly negative to positive values depending upon the basis set used. The MP2[16, 31] and coupled cluster methods[16], however, show negative electron affinity for both the NAB. The multi-reference CASPT2 method[16] gives very small value of adiabatic electron affinity almost close to zero for uracil and thymine. On the other hand, cytosine, guanine, and adenine give negative values in all the *ab-initio* methods.

Our EOMCC calculations also show that the AEA of all DNA and RNA nucleobases are negative in both aug-cc-pVDZ and aug-cc-pVTZ basis set. Figure 7 presents the EOMEA-CCSD natural orbitals for the NAB. It can be seen that the anions of first electron attached state of uracil and thymine are valence bound types, which is prominent from the localized nature of the natural orbitals. On the other hand, highly diffuse nature of the orbital corresponding to first electron attached state in cytosine, adenine, and guanine indicates the formation of a dipole bound anion. Now, the aug-cc-pVTZ is still not description of the states having a "dipole-bound" character. A more flexible basis set with additional diffuse might be necessary for a better description. However, such a big basis set EOMEA-CCSD calculation is beyond our present computation power. Moreover, use of a very diffuse basis will also give rise to problem of auto-ionizations due to resulting metastable state and special techniques like complex absorbing potential (CAP) should be used for their stabilization, which is outside the scope of the present study.

Here, all the electron affinity values for all the states are negative. Therefore, electron attached states are in reality quasi-bound or resonance state. The terms valence bound or dipole bound only indicates the nature of the interaction and they are true bound state in reality.

In Figure 8, we have plotted our EOMCC calculated AEA of all nucleobases. The plots show that the trend of VEA in aug-cc-pVTZ basis set i.e., G>U>T>C>A is conserved even in AEA. The uracil, thymine, cytosine and guanine show higher AEA values than that in adenine (see

Table 9). In all the cases, the value of electron affinity in adiabatic case increases (i.e., becomes less negative) from that in the vertical case.

Table 10 presents the deviation of first AEA from that of the VEA values in aug-cc-pVDZ and aug-cc-pVTZ basis sets. In aug-cc-pVDZ basis, the deviation roughly follows the trend in RMS deviations in geometry reported in Table 1. However, the agreement is at best qualitative and no quantitative inference can be drawn. For example, the large RMS deviation (0.49) in the case of thymine is not translated into the deviation of EA value from vertical to adiabatic state and the corresponding deviation in EA is of the same order of uracil which shows a considerably less RMS deviation of 0.079 in the geometry. Here, it should be noted that the large RMS deviation in thymine is mainly due to twisting of methyl group. It is well known that the change in dihedral angle has less effect on the energy than that caused by changing bond length and bond angle. Similar effect also persists in the case of cytosine, where the large RMSD value caused by twisting of the $NH_2$, does not translate into the large deviation of electron affinity from vertical to adiabatic and cytosine shows a deviation of 0.03 eV. On moving to aug-cc-pVTZ, the correlation between RMSD value and change in electron affinity becomes less straightforward. It is due to the fact that in aug-cc-pVTZ basis set, the relative deviations of EA from vertical to adiabatic state are too small to make a comparative analysis in the level of theory of used for their computation in the present study. One need to include partial triples and use larger basis sets in EOMEA-CC calculation to get the quantitatively accurate trend. However, such calculations are beyond our present computational power.

## 4. CONCLUSIONS

In this paper, we have studied vertical and adiabatic electron affinities of DNA and RNA nucleic acid bases using EOMEA-CCSD method. The EOMEA-CCSD method includes dynamic and non-dynamic correlation in balanced way and gives an opportunity to study multiple electron attached states in a single calculation. We have reported the first five electron attached states of all the NAB. Among the five states reported, only the first electron-attached state is accessible in the gas phase, and all the other states are in very high energy to hold the excess electron. The first vertical electron affinity of all five NAB are negative, i.e. the electron attached states are not bound, rather a resonance state. Among all the theoretical results available, EOMCC method gives the best agreement with experiment, and the experimental trend of their relative ordering is reproduced in the computed results. For Guanine, reliable experimental results are not available for comparison. EOMEA-CCSD method gives higher value of electron affinity for guanine (i.e. less negative) in contrast to the low electron affinity value (i.e. more negativity) reported in the previous theoretical studies.

An analysis of our B2PLYP/aug-cc-pVTZ optimized geometry shows that the structures of purine NAB thymine and uracil are more sensitive to the electron attachment, which is indicated by the large RMS deviation from their neutral geometry. On the other hand, the geometry of the pyrimidine NAB, adenine and guanine are less sensitive to the electron attachment. The cytosine falls in between this two classes. The trend in geometry change is also qualitatively reflected in the change of the electron affinity value from vertical to adiabatic state. Analysis of the natural orbitals obtained in the EOMEA-CCSD calculation shows that the first electron attached states obtained are valence-bound types for uracil and thymine, and dipole-bound type for cytosine, adenine and guanine, which explains the trend in the geometry change from the neutral molecules, on attachment of an extra electron. The AEA values are negative for all the NAB. The trend in relative VEA values of NAB is conserved in the AEA values. However, the AEA values are less negative than the corresponding VEA values.

As discussed previously, only the first electron-attached state is accessible in gas phase. However, previous theoretical studies[82-84] have shown that the presence of solvent molecules stabilizes the first electron attached state of NAB and thereby lowering down its energy. Similarly, the higher electron attached states of NAB can be accessible in solvent phase. Therefore, it will be interesting to study the effect of solvents on the electron affinities of different NAB using EOMEA-CCSD method. Work is currently under way towards that direction.

**ACKNOWLEDGEMENTS**
The authors acknowledge the grant from CSIR XIIth five year plan project on Multi-scale Simulations of Material and facilities of the Centre of Excellence in Scientific Computing at NCL. A.K.D thanks the Council of Scientific and Industrial Research (CSIR) for a Senior Research Fellowship. TS thanks the University Grant Commission (UGC) for a Junior Research

Fellowship. S.P. acknowledges the DST J. C. Bose Fellowship project and CSIR SSB grant towards completion of the work.

**Table 1. Structural changes of each nucleobases upon anion formation**[a]

| Nucleobases | Structural Parameters | Neutral State (N) | Anionic State (A) | Deviation (A-N) | RMSD Value |
|---|---|---|---|---|---|
| Adenine | Bond length(Å) | 1.352 (N5-C1) | 1.369 | +0.017 | |
| | Bond angle(deg) | 119.5(H2-N5-H3) | 117.2 | -2.3 | 0.032 |
| | Dihedral(deg) | -170.78 (H3-N5-C1-C2) | -163.42 | +7.36 | |
| Guanine | Bond length(Å) | 1.374(N5-C5) | 1.357 | -0.017 | |
| | Bond angle(deg) | 113.20(H3-N5-C5) | 115.44 | +2.24 | 0.041 |
| | Dihedral (deg) | 32.51(H2-N5-C5-N3) | 22.75 | -9.76 | |
| Uracil | Bond length(Å) | 1.345(C1-C2) | 1.415 | +0.070 | |
| | Bond angle(deg) | 121.96(N1-C1-C2) | 116.45 | -5.51 | 0.079 |
| | Dihedral (deg) | -0.001(C2-C1-N1-C4) | -16.16 | -16.16 | |
| Thymine | Bond length(Å) | 1.348(C2-C3) | 1.411 | +0.063 | |
| | Bond angle(deg) | 122.77(N2-C3-C2) | 117.84 | -4.93 | 0.49 |
| | Dihedral (deg) | -58.98 (H2-C5-C2-C1) | -126.38 | -67.40 | |
| Cytosine | Bond length(Å) | 1.218(C2-O1) | 1.227 | +0.009 | |
| | Bond angle(deg) | 117.27(H2-N3-C1) | 118.09 | +0.82 | 0.067 |
| | Dihedral (deg) | -167.99(H1-N3-C1-N1) | 171.07 | +20.94 | |

a: In each case only maximum deviation in bond length, bond angle and dihedral are tabulated.

**Table 2. Vertical electron affinities (eV) of DNA and RNA nucleobases obtained by different experimental methods**

| Reference | Method | Uracil | Thymine | Cytosine | Adenine | Guanine |
|---|---|---|---|---|---|---|
| Burrow et al[78]. | Expt. ETS | −0.22 | −0.29 | −0.32 | −0.54 | – |
| Desfrancois et al[79] | Cluster Solvation(RET) | -0.30 | -0.30 | -0.55 | -0.45 | – |
| Sangaranarayanan M et al[80] | Enthalpy of formation | -0.24 | -0.53 | -0.40 | -0.56 | -0.79 |

Table 3. Vertical electron affinities (eV) of DNA and RNA nucleobases obtained by different theoretical methods

| Reference | Method | Uracil | Thymine | Cytosine | Adenine | Guanine |
|---|---|---|---|---|---|---|
| Sevilla et al.[31] | Scaled Koopman/D95V | −0.11 | −0.32 | −0.40 | −0.74 | −1.23 |
| Reference[25-28] | B3LYP range | −1.09 to −0.11 | −1.05 to −0.28 | −1.42 to −0.31 | −1.57 to −0.34 | −2.07 to −0.08 |
| | MP2 /6-31G(d) | −1.77 | −1.85 | −1.97 | −2.54 | −2.82 |
| | PMP2 // MP2 / 6-31G (d) | −1.63 | −1.69 | −1.76 | −2.07 | −2.48 |
| | MP2/aug-cc-pVDZ | −0.69 | −0.73 | −0.91 | −1.42 | −1.57 |
| | PMP2 // MP2 / aug-cc-pVDZ | −0.56 | −0.58 | −0.73 | −0.99 | −1.30 |
| | CCSD// CCSD/ aug-cc-pVDZ | −0.63 | −0.65 | −0.77 | --- | --- |
| Serrano-Andrés et al[16] | CCSD (T) // CCSD/aug-cc-pVDZ | −0.64 | −0.65 | −0.79 | --- | --- |
| | CASPT2 // CASSCF/ cc-pVDZ | −1.42 | −1.44 | −1.49 | −1.65 | −2.14 |
| | CASPT2 //CASSCF/ ANO-L 431/21 | −0.68 | −0.69 | −0.76 | −1.06 | −1.30 |
| | CASPT2/ANO-L 4321/ 321// CASSCF/ ANO-L 431/21 | −0.49 | −0.45 | −0.59 | −0.74 | −0.94 |
| | CASPT2 IPEA /ANO-L 4321/ 321// CASSCF/ ANO-L 431/ 21 | −0.61 | −0.60 | −0.69 | −0.91 | −1.14 |
| Present Work | EOMCC/ aug-cc-pVDZ//B2PLYP/aug-cc-pVTZ | -0.23 | -0.28 | -0.27 | -0.44 | -0.23 |
| Present Work | EOMCC/ aug-cc-pVTZ[a]//B2PLYP/aug-cc-pVTZ | -0.20 | -0.24 | -0.24 | -0.40 | -0.19 |

a : for hydrogen atoms aug-cc-pVDZ basis is used

**Table 4. Vertical electron affinities (eV) for five states of DNA and RNA nucleobases in EOMEA-CCSD method.**

| State | Basis Set | Uracil | Thymine | Cytosine(C1) | Adenine | Guanine(G9k) |
|---|---|---|---|---|---|---|
| State 0 | aug-cc-pVDZ | -0.23 | -0.28 | -0.27 | -0.44 | -0.23 |
|  | aug-cc-pVTZ | -0.20 | -0.24 | -0.24 | -0.40 | -0.19 |
| State 1 | aug-cc-pVDZ | -0.72 | -0.73 | -0.55 | -0.76 | -0.50 |
|  | aug-cc-pVTZ | -0.60 | -0.62 | -0.51 | -0.70 | -0.45 |
| State 2 | aug-cc-pVDZ | -0.90 | -0.79 | -0.92 | -1.00 | -0.94 |
|  | aug-cc-pVTZ | -0.84 | -0.74 | -0.79 | -0.94 | -0.88 |
| State 3 | aug-cc-pVDZ | -1.04 | -0.99 | -0.99 | -1.21 | -1.21 |
|  | aug-cc-pVTZ | -0.99 | -0.93 | -0.94 | -1.07 | -1.16 |
| State 4 | aug-cc-pVDZ | -1.47 | -1.40 | -1.43 | -1.28 | -1.41 |
|  | aug-cc-pVTZ | -1.38 | -1.29 | -1.36 | -1.21 | -1.26 |

Table 5 : Vertical electron affinities (eV) for first electron attached states of cytosine in EOMEA-CCSD method.

| Basis Set | C2a | C3a | C3b | C2b | C1 |
|---|---|---|---|---|---|
| aug-cc-pVDZ | -0.38 | -0.28 | -0.35 | -0.43 | -0.27 |
| aug-cc-pVTZ | -0.34 | -0.25 | -0.32 | -0.38 | -0.24 |

Table 6 : Vertical electron affinities (eV) for first electron attached states of different isomers of guanine in EOMEA-CCSD method.

| Basis Set | G7k | G7ES | G9EA | G9ES | G9K |
|---|---|---|---|---|---|
| aug-cc-pVDZ | -0.32 | -0.39 | -0.41 | -0.48 | -0.23 |
| aug-cc-pVTZ | -0.23 | -0.35 | -0.37 | -0.43 | -0.19 |

**Table 7: Relative ordering of vertical electron affinity values (eV) of NAB for first five electron attached state obtained in EOMEA-CCSD method**

| States | Basis | Trend |
| --- | --- | --- |
| State 0 | aug-cc-pVDZ | G~U>C>T>A |
|  | aug-cc-pVTZ | G>U>C~T>A |
| State 1 | aug-cc-pVDZ | G>C>U>T>A |
|  | aug-cc-pVTZ | G>C>U>T>A |
| State 2 | aug-cc-pVDZ | T>U>C>G>A |
|  | aug-cc-pVTZ | T>C>U>G>A |
| State 3 | aug-cc-pVDZ | T~C>U>A~G |
|  | aug-cc-pVTZ | T>C>U>A>G |
| State 4 | aug-cc-pVDZ | A>T>G>C>U |
|  | aug-cc-pVTZ | A>G>T>C>U |

**Table 8. Experimental values[a] of adiabatic electron affinity (eV) of Uracil and Thymine**

| References | Uracil | Thymine |
|---|---|---|
| Wentworth et al.[86] | 0.75 | 0.65 |
| Wentworth et al.[87, 88] | 0.80 | 0.79 |
| Weinkauf et al.[89] | 0.15 ± 0.12 | 0.12 ± 0.12 |
| Schermann et al.[96] | > 30 − 60 and < 93 | – |
| Desfrancois et al[79] | ≈ 0 | ≈ 0 |
| Sanche et al.[97] | – | > 0 |

a : Taken from ref 85

**Table 9. Adiabatic electron affinities (eV) of DNA and RNA nucleobases obtained by different theoretical methods.**

| Reference | Method | Uracil | Thymine | Cytosine | Adenine | Guanine |
|---|---|---|---|---|---|---|
| Sevilla et al.[31] | Scaled Koopman/D95V | 0.4 | 0.3 | 0.2 | -0.3 | -0.7 |
| Sevilla et al[31] | Scaled MP2/6-31+G(d)//MP2/6-31G* | -0.25 | -0.30 | -0.46 | -1.19 | -0.75 |
| Sevilla et al.[27] | B3LYP/ 6-31G $d$ | -0.52 | -0.49 | -0.69 | -1.18 | -1.51 |
| Sevilla et al.[27] | B3LYP/ 6-311+ +G (2$d$ , $p$) | 0.20 | 0.22 | -0.05 | -0.30 | -0.01 |
| Walch[90] | B3LYP/ 6-31+ +G (Ryd ) | --- | 0.34 | 0.20 | 0.08 | 0.25 |
| Boyd et al[92-94] | B3LYP/6-311G(2df,p)//B3LYP/6-31G(d,p) | -0.4 | -0.64 | --- | -0.9 | -0.69 |
| Russo et al.[95] | B3LYP/6-311++G//B3LYP/6-311++G** | 0.21 | 0.18 | 0.006 | -0.26 | -0.004 |
| Schaefer et al.[91] | B3LYP/TZ2P++//B3LYP/DZP++ | 0.19 | 0.16 | -0.02 | -0.17 | 0.07 |
| Serrano-Andrés et al[16] | MP2 /6-31G $d$ | -1.16 | -1.20 | -1.31 | -1.98 | -1.63 |
| | PMP2 //MP2 /6-31G( $d$) | -1.06 | -1.09 | -1.17 | -1.81 | -1.55 |
| | MP2/aug-cc-pVDZ | -0.21 | -0.26 | -0.40 | -1.06 | -0.71 |
| | PMP2 // MP2/ aug-cc-pVDZ | -0.09 | -0.14 | -0.25 | -0.88 | -0.63 |
| | CCSD/aug-cc-pVDZ | -0.07 | -0.12 | -0.18 | -0.90 | -0.50 |
| | CCSD (T) / CCSD/aug-cc-pVDZ | -0.05 | -0.09 | -0.17 | -0.84 | -0.44 |
| | CASPT2 //CASSCF/ cc-pVDZ | -0.69 | -0.74 | -0.77 | -1.02 | -1.00 |
| | CASPT2 /CASSCF/ANO-L 431/21 | -0.03 | 0.06 | -0.03 | -0.54 | -0.31 |
| | CASPT2/ANO-L 4321/321//CASSCF/ANO-L 431/ 21 | 0.10 | 0.19 | 0.08 | -0.39 | -0.20 |
| | CASPT2 IPEA /ANO-L 4321/ 321//CASSCF/ANO-L 431/21 | -0.01 | 0.05 | -0.04 | -0.57 | -0.35 |
| | CASPT2 IPEA /ANO-L 4321/ 321// CCSD/aug-cc-pVDZ | 0.03 | 0.02 | -0.10 | -0.72 | -0.44 |
| Present work | EOMCC/ aug-cc-pVDZ//B2PLYP/aug-cc-pVTZ+ZPE | -0.13 | -0.19 | -0.24 | -0.41 | -0.15 |
| Present work | EOMCC/ aug-cc-pVTZ[a]//B2PLYP/aug-cc-pVTZ+ZPE | -0.15 | -0.2 | -0.21 | -0.38 | -0.12 |

a: for hydrogen atoms aug-cc-pVDZ basis is used

**Table 10. Deviation of AEA from the VEA values in aug-cc-pVDZ and aug-cc-pVTZ basis set (in eV)**

| Basis set   | Uracil | Thymine | Cytosine | Adenine | Guanine |
|-------------|--------|---------|----------|---------|---------|
| aug-cc-pVDZ | 0.10   | 0.09    | 0.03     | 0.03    | 0.07    |
| aug-cc-pVTZ | 0.05   | 0.04    | 0.03     | 0.02    | 0.07    |

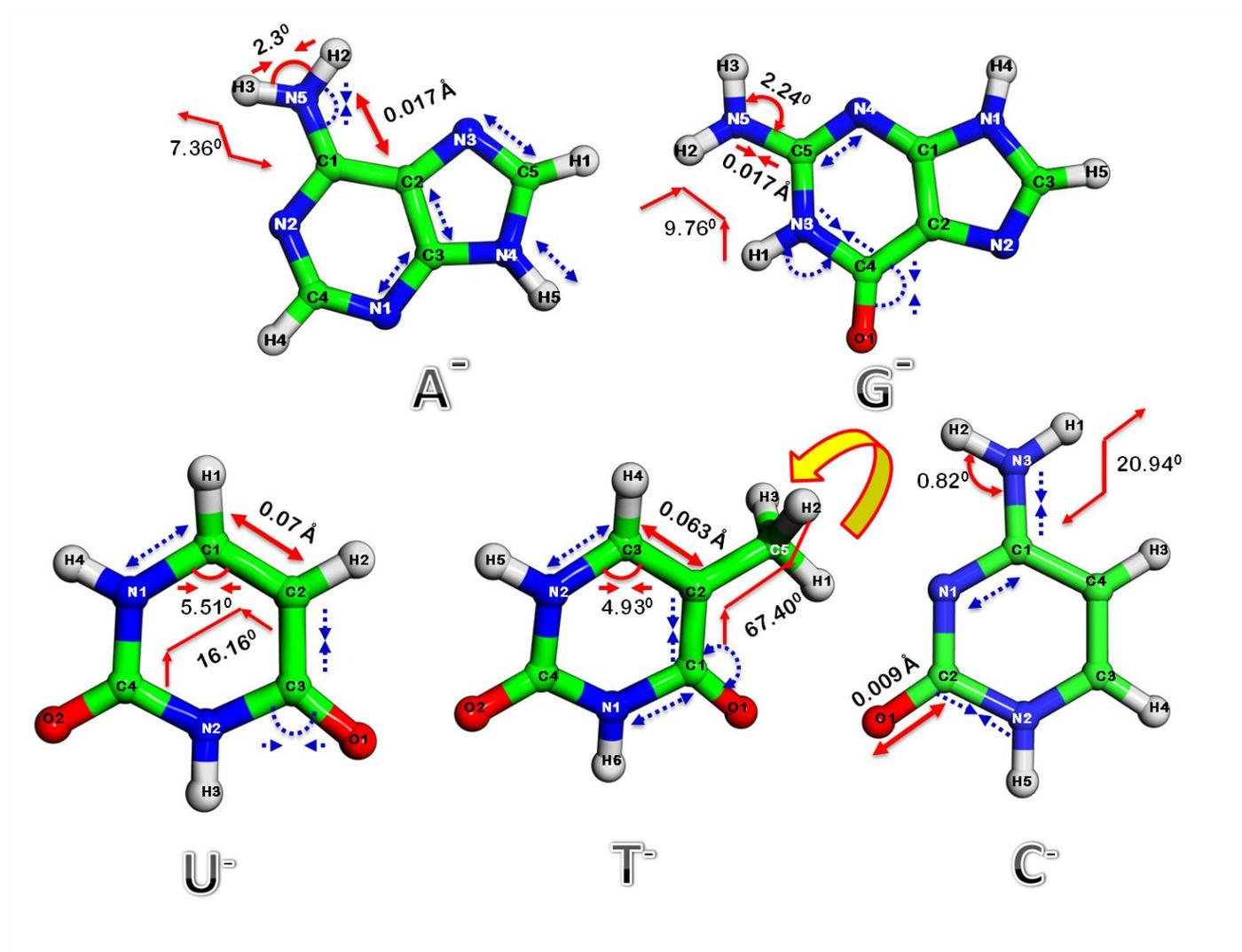

Figure 1. Optimised geometries of anions of nucleobases obtained by B2PLYP/aug-cc-pVTZ level of theory, along with maximum deviations in structural parameters (solid red line), other major structural changes in bond angles and bond lengths are shown by dotted (blue) line. Outward arrows signify increment of corresponding parameter in anionic form and vice-versa.

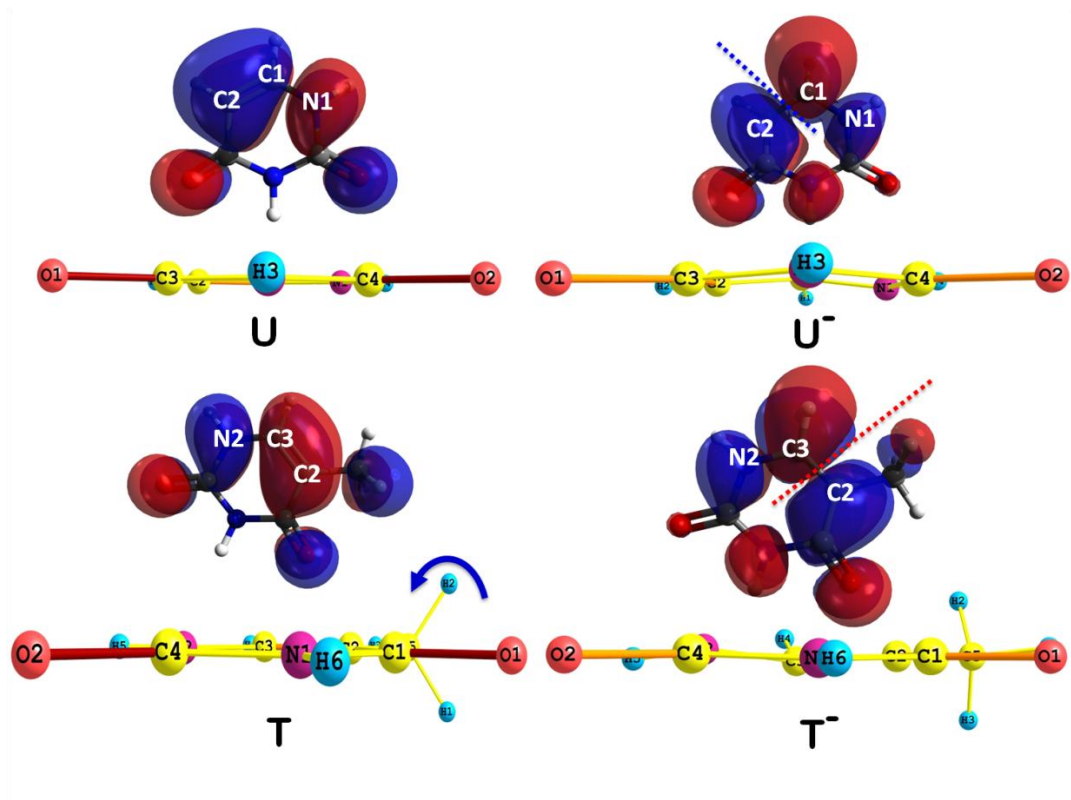

Figure 2: Comparison of HOMO of uracil and thymine with SOMO of their anions shows introduction of new anti-bonding interaction in anionic form, which can reasonably explain major structural changes in anions of uracil of thymine. Planner side view of both shows that each molecule has lost its ring planarity after electron attachment.

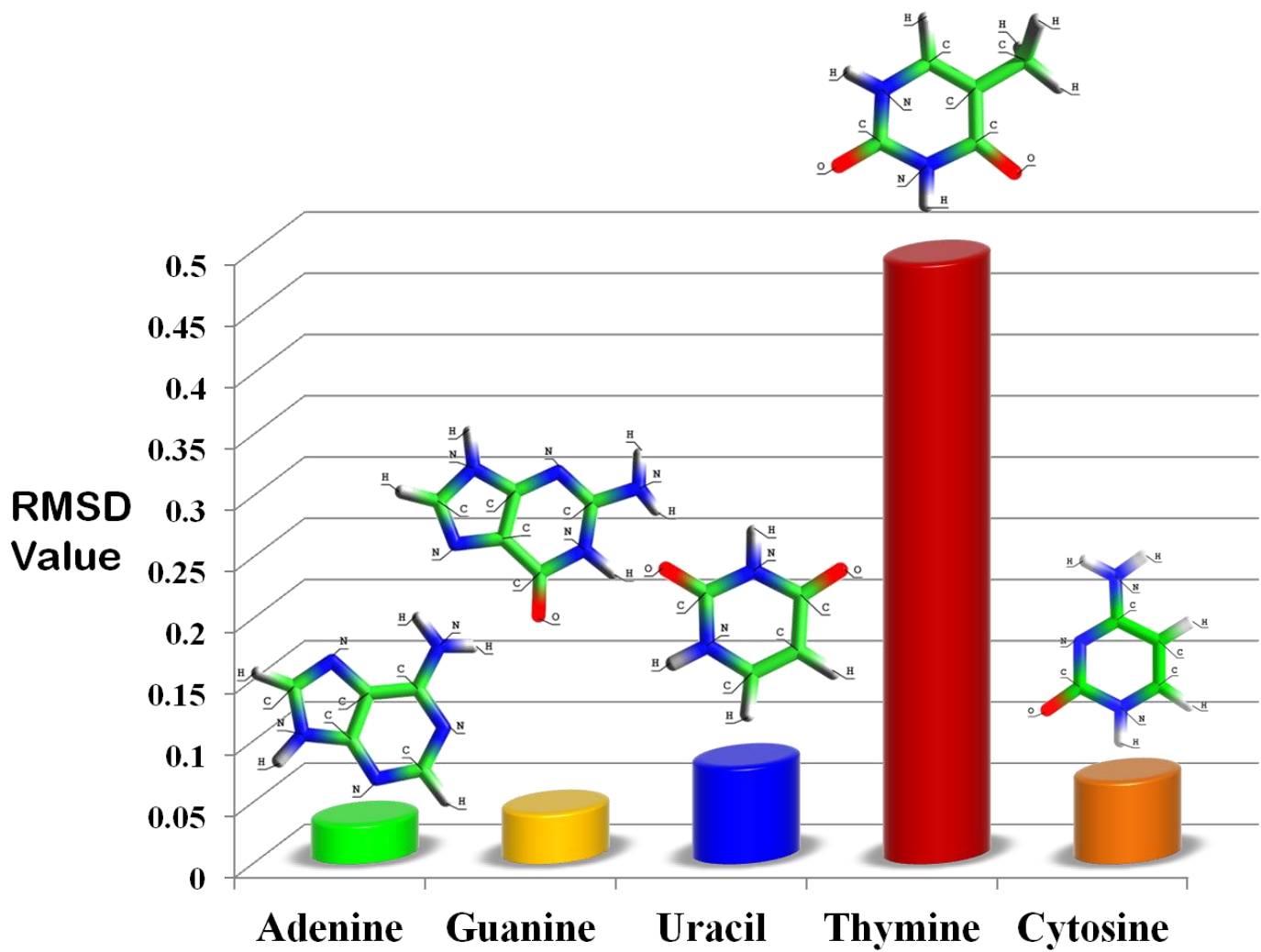

Figure 3: RMSD plot of NABs showing thymine having the highest RMSD value among all NAB's. Principle reason of this is the rotation of $-CH_3$ group in the anion of thymine.

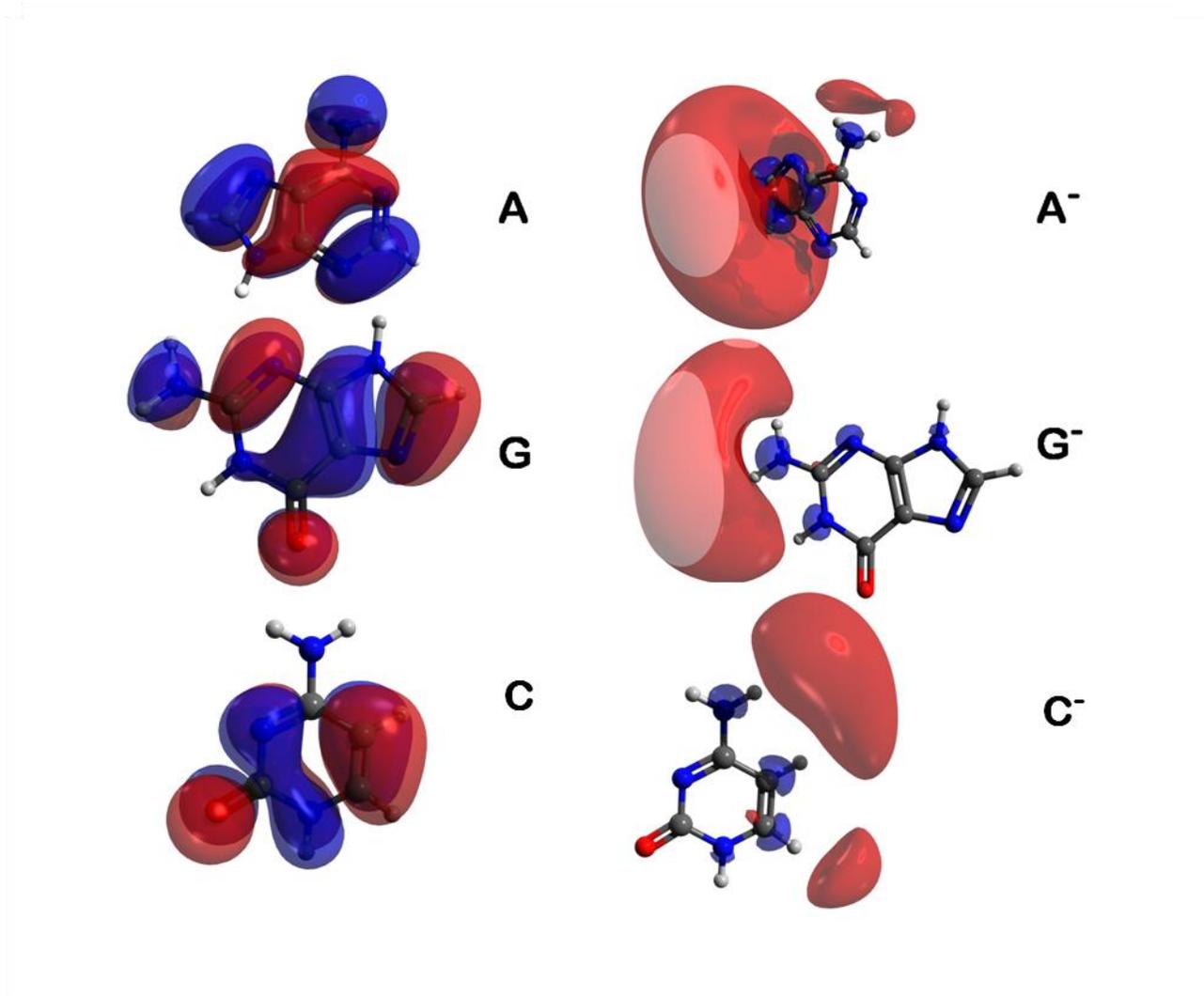

Figure 4: Qualitative comparison of HOMO of Adenine Guanine and Cytosine and SOMO of their anions obtained via B2PLYP/aug-cc-pVTZ level of theory. SOMO of A,G and C anions are found to be dipole bound in nature.

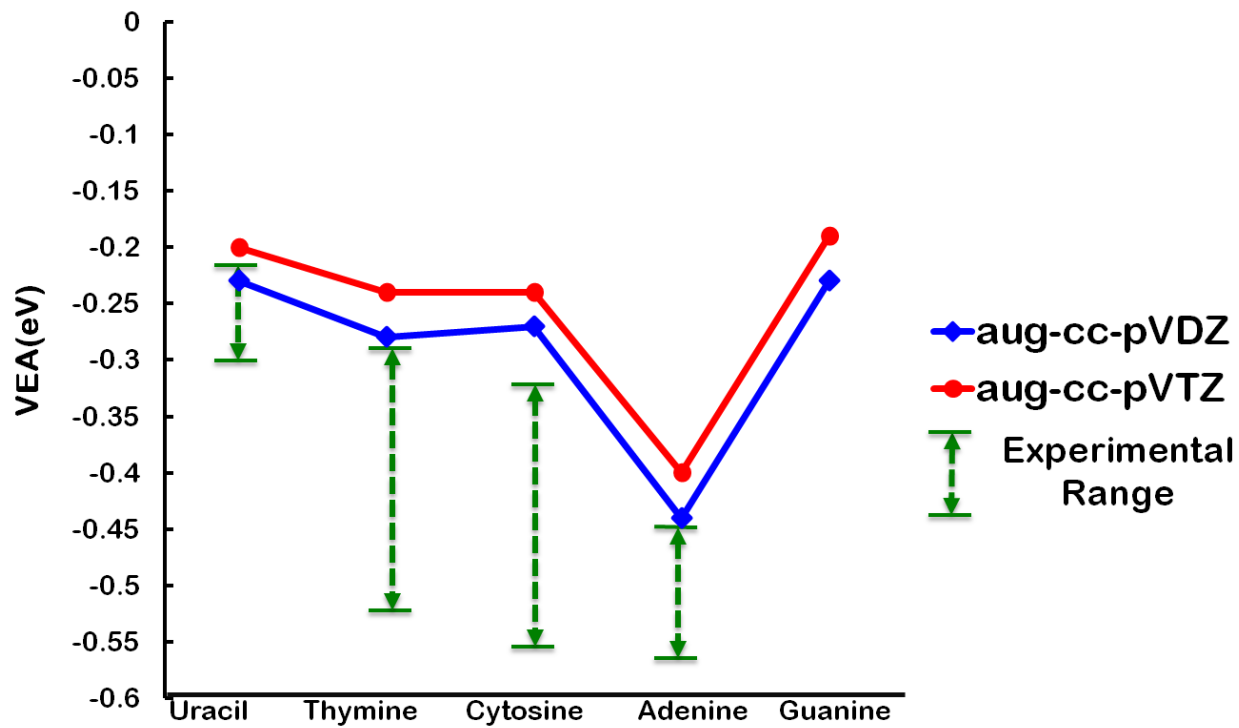

Figure 5 : The comparison between EOMCC and experimental valence VEA values. Experimental range of guanine is not presented due to unavailability of experimental results.

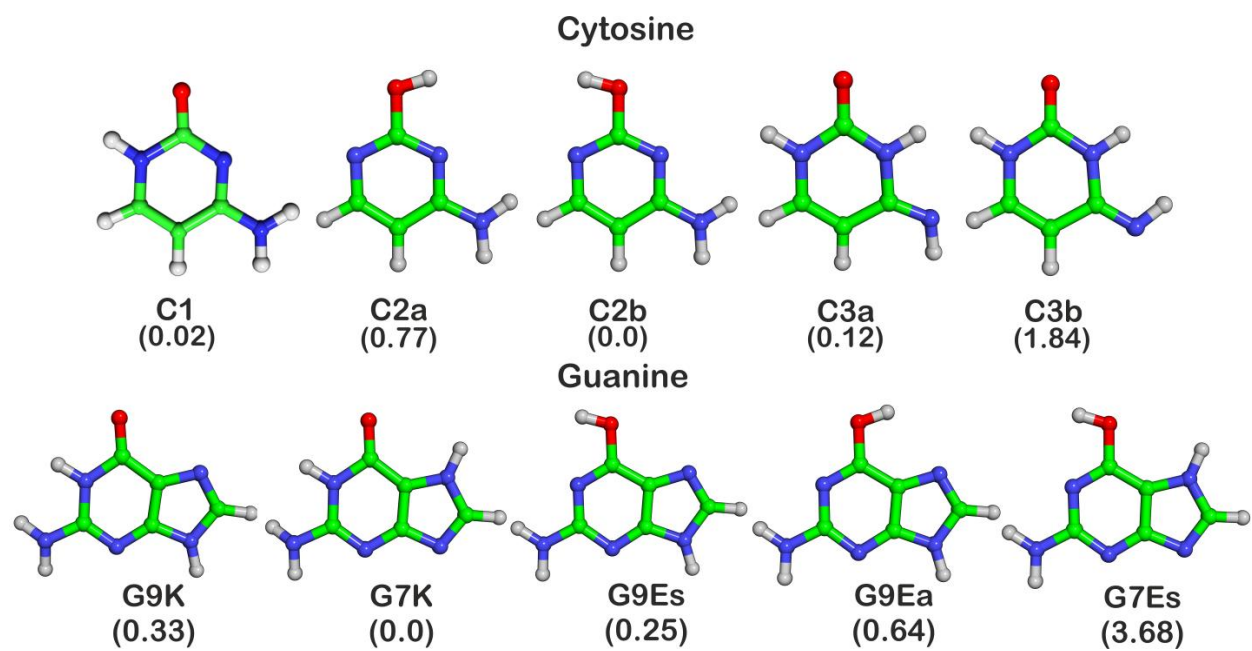

Figure 6 : Different isomers of Cytosine and Guanine. Relative energy in Kcal/mol is provided in the bracket

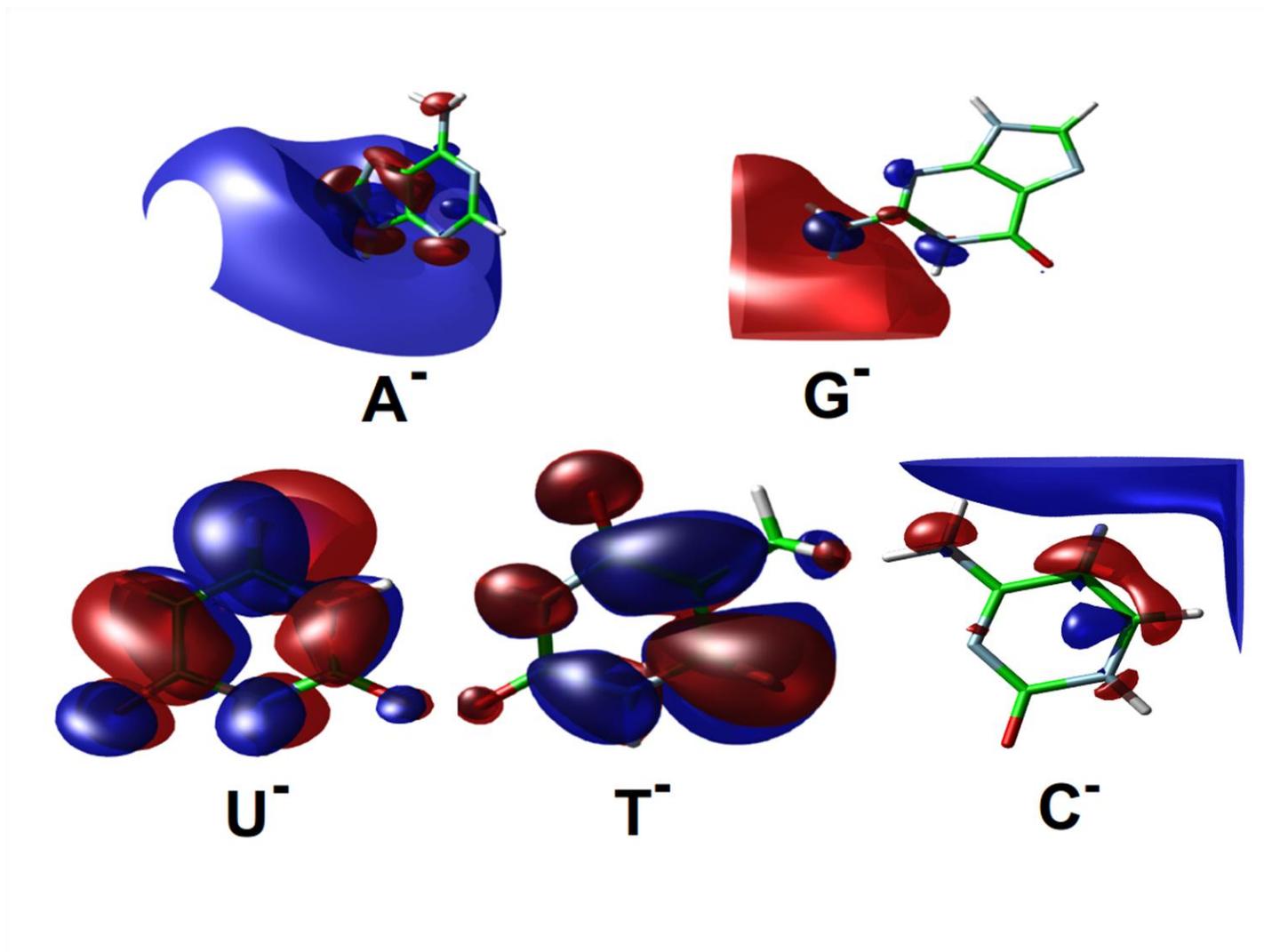

Figure 7: Natural orbitals of anions (SOMO) of five nucleobases obtained in EOMEA-CCSD level of theory.

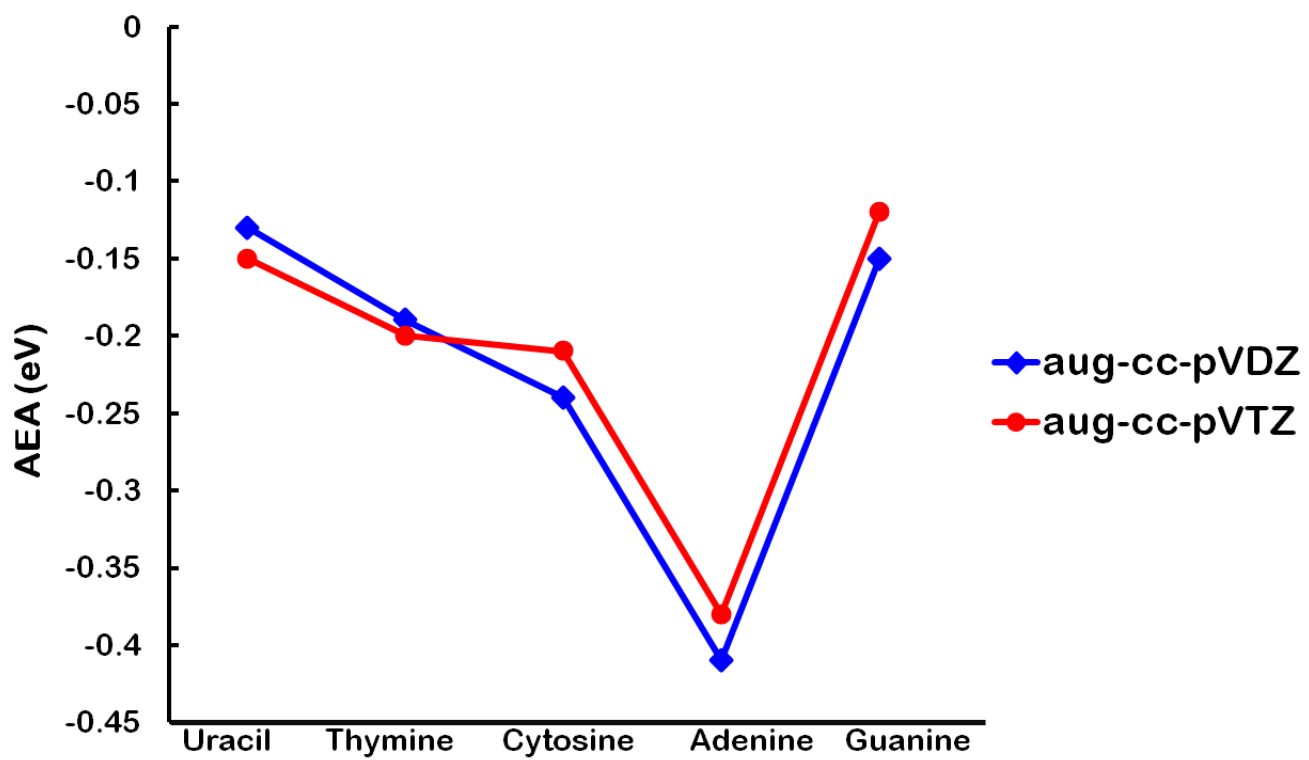

Figure 8: Trends of adiabatic electron affinities of nucleobases obtained by EOMCC method in both the basis sets.